\def\jnl@style{\it}
\def\aaref@jnl#1{{\jnl@style#1}}

\def\aaref@jnl#1{{\jnl@style#1}}

\def\aj{\aaref@jnl{AJ}}                   
\def\araa{\aaref@jnl{ARA\&A}}             
\def\apj{\aaref@jnl{ApJ}}                 
\def\apjl{\aaref@jnl{ApJ}}                
\def\apjs{\aaref@jnl{ApJS}}               
\def\ao{\aaref@jnl{Appl.~Opt.}}           
\def\apss{\aaref@jnl{Ap\&SS}}             
\def\aap{\aaref@jnl{A\&A}}                
\def\aapr{\aaref@jnl{A\&A~Rev.}}          
\def\aaps{\aaref@jnl{A\&AS}}              
\def\azh{\aaref@jnl{AZh}}                 
\def\baas{\aaref@jnl{BAAS}}               
\def\jrasc{\aaref@jnl{JRASC}}             
\def\memras{\aaref@jnl{MmRAS}}            
\def\mnras{\aaref@jnl{MNRAS}}             
\def\pra{\aaref@jnl{Phys.~Rev.~A}}        
\def\prb{\aaref@jnl{Phys.~Rev.~B}}        
\def\prc{\aaref@jnl{Phys.~Rev.~C}}        
\def\prd{\aaref@jnl{Phys.~Rev.~D}}        
\def\pre{\aaref@jnl{Phys.~Rev.~E}}        
\def\prl{\aaref@jnl{Phys.~Rev.~Lett.}}    
\def\pasp{\aaref@jnl{PASP}}               
\def\pasj{\aaref@jnl{PASJ}}               
\def\qjras{\aaref@jnl{QJRAS}}             
\def\skytel{\aaref@jnl{S\&T}}             
\def\solphys{\aaref@jnl{Sol.~Phys.}}      
\def\sovast{\aaref@jnl{Soviet~Ast.}}      
\def\ssr{\aaref@jnl{Space~Sci.~Rev.}}     
\def\zap{\aaref@jnl{ZAp}}                 
\def\nat{\aaref@jnl{Nature}}              
\def\iaucirc{\aaref@jnl{IAU~Circ.}}       
\def\aplett{\aaref@jnl{Astrophys.~Lett.}} 
\def\apspr{\aaref@jnl{Astrophys.~Space~Phys.~Res.}}
\def\bain{\aaref@jnl{Bull.~Astron.~Inst.~Netherlands}} 
\def\fcp{\aaref@jnl{Fund.~Cosmic~Phys.}}  
\def\gca{\aaref@jnl{Geochim.~Cosmochim.~Acta}}   
\def\grl{\aaref@jnl{Geophys.~Res.~Lett.}} 
\def\jcp{\aaref@jnl{J.~Chem.~Phys.}}      
\def\jgr{\aaref@jnl{J.~Geophys.~Res.}}    
\def\jqsrt{\aaref@jnl{J.~Quant.~Spec.~Radiat.~Transf.}}
\def\memsai{\aaref@jnl{Mem.~Soc.~Astron.~Italiana}}
\def\nphysa{\aaref@jnl{Nucl.~Phys.~A}}   
\def\physrep{\aaref@jnl{Phys.~Rep.}}   
\def\physscr{\aaref@jnl{Phys.~Scr}}   
\def\planss{\aaref@jnl{Planet.~Space~Sci.}}   
\def\procspie{\aaref@jnl{Proc.~SPIE}}   

\newcommand{\Vtot}{V_{\mathrm{tot}}}
\newcommand{\Vdisk}{V_{\mathrm{disk}}}
\newcommand{\Vdata}{V_{\mathrm{data}}}
\newcommand{\AB}{\mathrm{AB}}
\newcommand{\BC}{\mathrm{BC}}
\newcommand{\CA}{\mathrm{CA}}
\newcommand{\data}{\mathrm{data}}
\newcommand{\turb}{\mathrm{turb}}
\newcommand{\bornes}{_{\lambda - \frac{\Delta \lambda}{2}}^{\lambda + \frac{\Delta \lambda}{2}}}

\documentclass[a4paper]{spie}  
\usepackage[]{graphicx}
\title{Intricate visibility effects from resolved emission of young stellar objects: the case of MWC158 observed with the VLTI} 

\author{Kluska J.\supit{a}, Malbet F.\supit{a}, Berger J.-P.\supit{a,b}, Lazareff B.\supit{a}, Le Bouquin J.-B.\supit{a}, Benisty M.\supit{a}, Menard F.\supit{a}, Pinte C.\supit{a}, Millan-Gabet R.\supit{c}, Traub W.\supit{d}.
\skiplinehalf
\supit{a}Institut de Plan\'etologie et d Astrophysique de Grenoble (UMR 5274)
BP 53
F-38041 GRENOBLE Cedex 9 , France; \\
\supit{b}ESO, Santiago Office,
Alonso de Cordova 3107,
Vitacura, Casilla 19001, Santiago de Chile, Chile; \\
\supit{c}California Institute of Technology
770 S. Wilson Ave. MS 100-22
Pasadena, CA 91125, USA; \\
\supit{d}Jet Propulsion Laboratory, California Institute of Technology, M/S 321-100, 4800 Oak Grove Drive Pasadena, CA 91109, USA
}

\authorinfo{Further author information: (Send correspondence to J. Kluska)\\ 
J. Kluska: E-mail: jacques.kluska@obs.ujf-grenoble.fr, Telephone:  (+33) 4 56 52 09 66}

\begin{document} 
  \maketitle 

\begin{abstract}
In the course of our VLTI young stellar object PIONIER imaging program,
we have identified a strong visibility chromatic dependency that
appeared in certain sources. This effect, rising value of visibilities
with decreasing wavelengths over one base, is also present in previous published and
archival AMBER data. For Herbig AeBe stars, the H band is generally located at the
transition between the star and the disk predominance in flux for Herbig
AeBe stars. We believe that this phenomenon is responsible for the
visibility rise effect. We present a method to correct the visibilities
from this effect in order to allow "gray" image reconstruction software,
like $\tt{Mira}$, to be used. In parallel we probe the interest of carrying an
image reconstruction in each spectral channel and then combine them to
obtain the final broadband one. As an illustration we apply these
imaging methods to MWC158, a (possibly Herbig) B[e] star intensively
observed with PIONIER. Finally, we compare our result with a parametric
model fitted onto the data.\end{abstract}



\section{INTRODUCTION}
\label{sec:intro}  

The processes that lead to the formation of exoplanets are important to understand. Stars form after a collapse of a giant cloud of dust and gas. After a million year, a protoplanetary disk is forming around the star, believed to be the birthplace of planets. 

A young star is surrounded by an active environment with which it interacts. Accretion disks\cite{Monnier}, inner gaseous disks\cite{gas1,gas2,gas3}, infalling envelop renmands, winds\cite{winds1,winds2,Malbet,Tatulli} and jets\cite{jets1,jets2} are the main components of such environments. 
The complexity of physical phenomenon at play requires direct observation at the astronomical unit (A.U.) scale.
Optical interferometry is able to bring such informations, because it can observe both in the near infrared, where the hot dust and hot gas nearby the star are emitting, and resolve the first A.U., which correspond to milliarcsecond scale at the distance of star formation regions.

\begin{figure}[!t]
   \begin{center}
   \begin{tabular}{cc}
   \includegraphics[height=7cm]{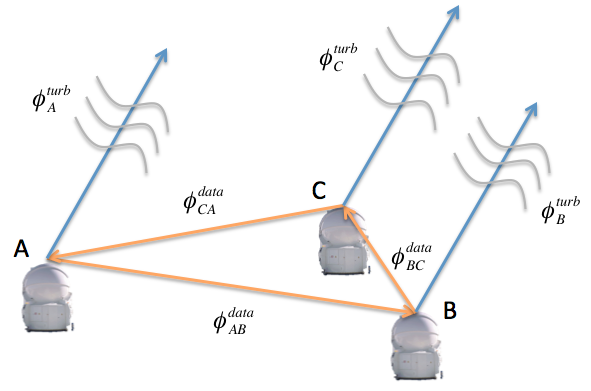}
   \end{tabular}
   \end{center}
   \caption[] 
   { \label{fig:CP} 
   Sketch representing three Auxiliary telescopes from the VLT Interferometer. Each telescope have a phase coming from the atmospheric turbulence. This phase will blur the astrophysical phase. In order to retrieve the physical information, we will compute the closure phase (Eq. \ref{eqn:closurephase} and Eq. \ref{eqn:closurephase2}). }
   \end{figure}

Interferometry consists in combining the light of 2 or more telescopes in order to measure the complex degree of coherence.
For that purpose, the interferometer measures interference fringes. The amplitude of the fringes yields the norm, and its position the phase of a complex quantity called visibility $V(u,v)$. Thanks to the van Cittert-Zernicke theorem we know that the Fourier transform of the visibilities in the Fourier Plan $(u,v)$ gives us the intensity distribution $I(x,y)$ of the source. Unfortunately, in the near infrared (NIR) the atmosphere blurs the phases of the visibilities. In practice, there are two interferometric measurements : the squared amplitude of the visibilities $V^2$ is the first one.
The second measurement of optical interferometry is the bispectra\cite{Jennison} and particularly their phase. A bispectrum is a product of visibilities on three different baselines forming a triangle. If we consider three telescopes A, B and C (see Fig.\ref{fig:CP}) the bispectrum will be :
\begin{equation}
\label{eqn:bispectrum}
bispectrum : V_{\AB} V_{\BC} V_{\CA}
\end{equation}
The phase of the bispectrum is : 
\begin{equation}
\label{eqn:bispectrumphase}
V_{\AB} V_{\BC} V_{\CA} = |V_{\AB}| e^{\phi_{\AB}} |V_{\BC}| e^{\phi_{\BC}} |V_{\CA}| e^{\phi_{\CA}}
\end{equation}
\begin{equation}
\label{eqn:closure}
\phi_{bispectrum} =\phi_{\AB} + \phi_{\BC} + \phi_{\CA}
\end{equation}
Since we do not have the phases because of the atmosphere turbulence, each telescope have residual phase. However, the bispectrum phase is astrophysical because this quantity cancels the turbulence effects : 
\begin{equation}
\label{eqn:closurephase}
\phi_{bispectrum} = \phi_{\AB}^{\data} + \phi_B^{\turb} - \phi_A^{\turb} + \phi_{\BC}^{\data} + \phi_C^{\turb} - \phi_B^{\turb} + \phi_{\CA}^{\data} + \phi_A^{\turb} - \phi_C^{\turb} \\
\end{equation}
and finally :
\begin{equation}
\label{eqn:closurephase2}
\phi_{\mathrm{bispectrum}} = \phi_{\AB}^{\data} + \phi_{\BC}^{\data}  + \phi_{\CA}^{\data} 
\end{equation}

\begin{figure}[!t]
   \begin{center}
   \begin{tabular}{cc}
   \includegraphics[height=5cm]{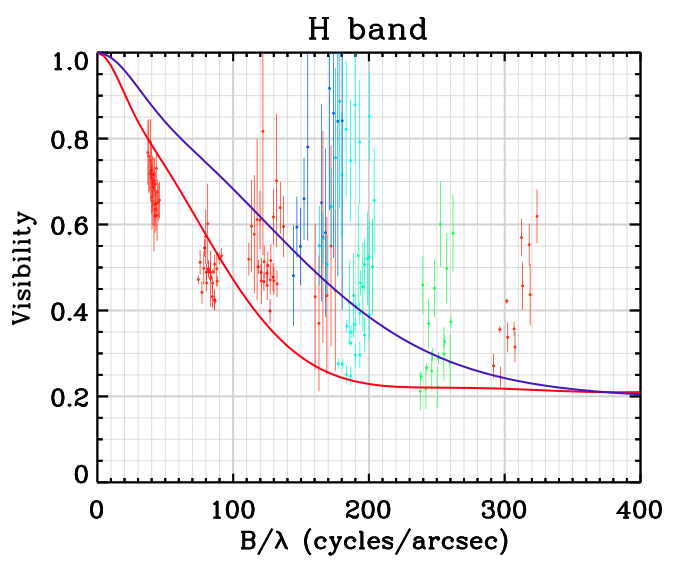}
    \includegraphics[height=5cm]{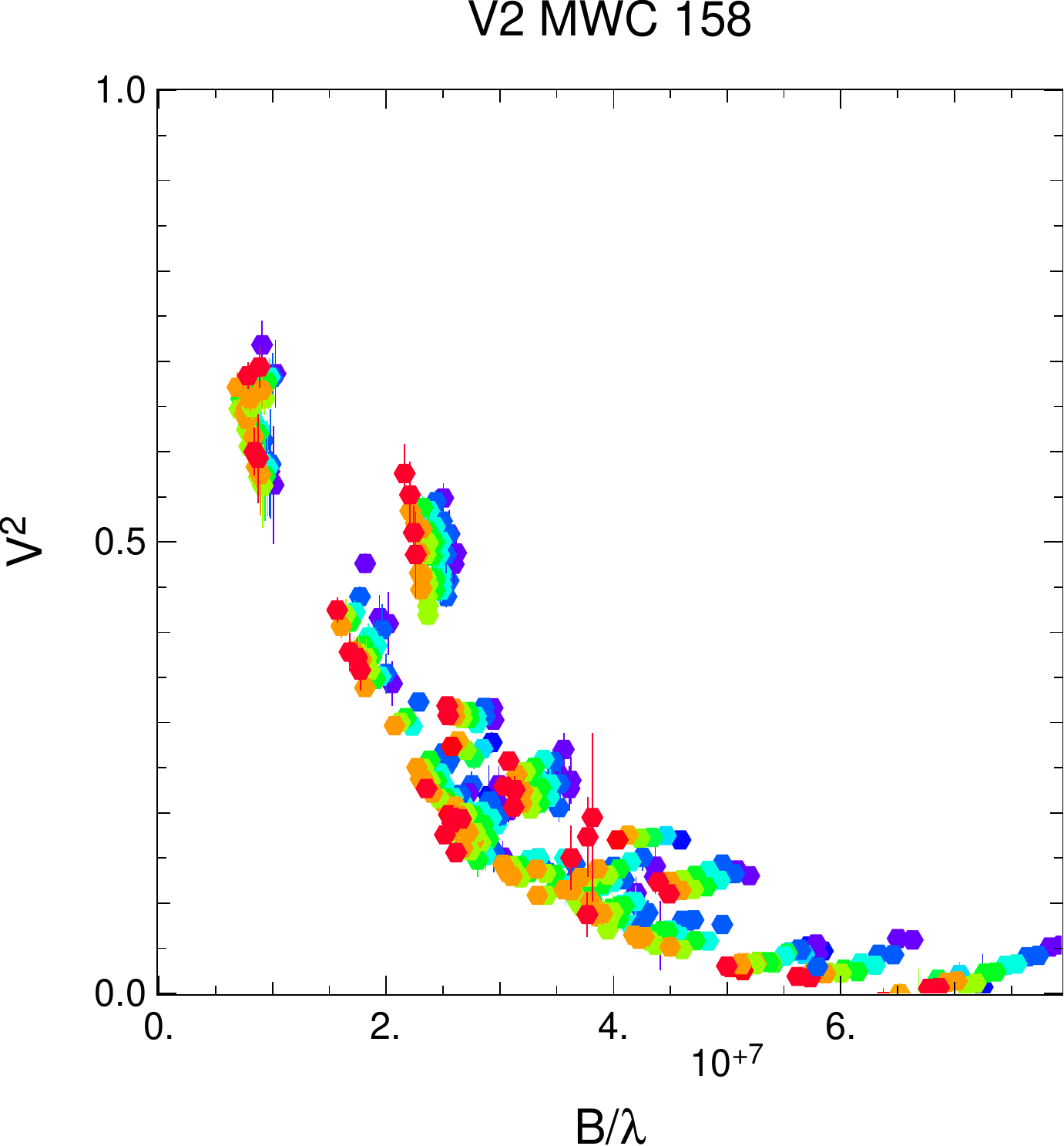}
     \includegraphics[height=5cm]{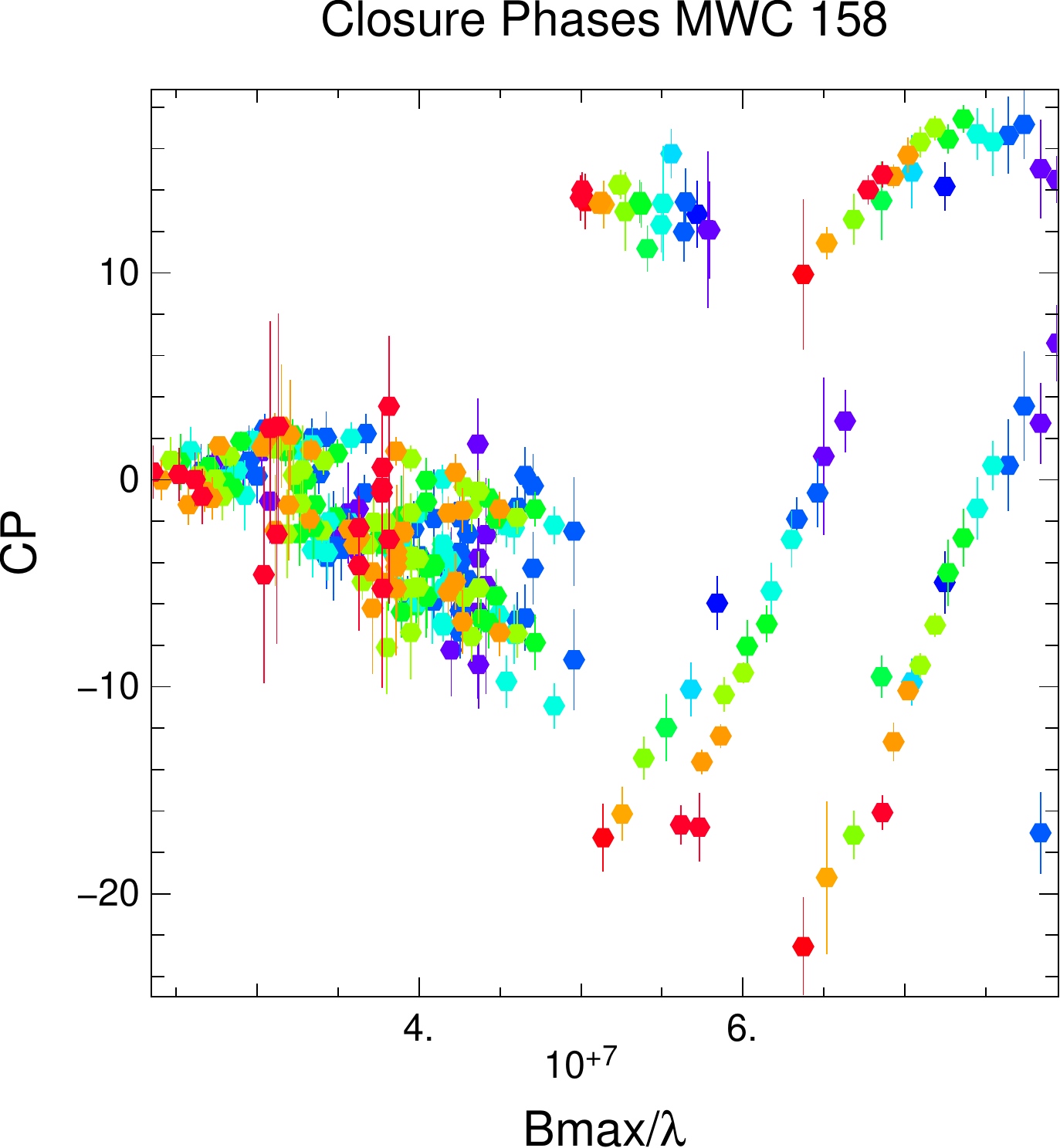}
   \end{tabular}
   \end{center}
   \caption[Interferometric data on MWC158] 
   { \label{fig:data} 
Data on MWC 158. Left : AMBER\cite{Petrov} data from ref. \citenum{Borges2011}. Center : PIONIER\cite{JBLB} squared visiblities. Right : PIONIER closure phases. For PIONIER data the color is in function of wavelength (blue : short wavelengths, red : long wavelengths).}
   \end{figure}

For each pair of telescopes, one can make multiple measurements (in the ($u,v$) plane) thanks to the number of spectral channels (since the spatial frequency ($f = \sqrt{u^2 + v^2}$) is equal to the baseline length ($B$) divided by the wavelength ($\lambda$) : $f=B/\lambda$). This is called the spectral super synthesis. We noticed in several datasets that the visibility is higher at short wavelengths. If we plot the squared visibilities $V^2$ in function of $B/\lambda$, we can see (Fig. \ref{fig:data}) that the rising curve of visibilities per base is not fitting the general trend of the data for different baselines. 
First, it was seen in AMBER\cite{Petrov} data, but it was considered as an instrumental defect. Now, the same effect has been observed with PIONIER \cite{JBLB} .
We try to explain this effect astrophysically, claiming that the image of the object is varying through the different spectral channels inside the same spectral band, and we propose three techniques in order to take it into account and to be able to reconstruct images.

These methods will be applied to an astrophysical object. They are useful to analyze MWC158 (also known as HD50138). This star is a Be star known to have the B[e] phenomenon and presents a strong variability\cite{Huts,Andrillat,Pogodin,Borges2009} which complexify the evolutionary stage identification of the source. Its distance is poorly constrain ($d = 500pc \pm 150pc$\cite{vl}). 

In section \ref{polychromatism} we will describe the chromatic effects in the visibilities and the section \ref{methods} will show the different methods to deal with them. Finally we will apply them to the astrophysical case of MWC 158 in the section \ref{mwc158}.

\section{CHROMATISM}
\label{polychromatism}
Since interferometric instruments with spectral dispersion exist, we need to take into account the flux variations with the wavelength in order to correctly analyze the data and have access to the spectral super synthesis. In the case of Young Stellar Objects (YSOs), we noticed that the visibilities have a strong spectral dependence such as the geometrical shape of the object could not explain it.  
For Herbig AeBe star, the chromatic effect explained in the section \ref{sec:intro} exists typically for the Near Infrared interferometry.
In the following, we explore the possibility that this effect is caused by a different spectral index between the central star and its surrounding media.

\subsection{Modeling the effect}
In order to confirm that, we made a simple model with a central star and its dusty disk.
\subsubsection{The star}
In our model, the star is considered to be unresolved. This hypothesis is justified for the young objects we are looking at. If we suppose a young Herbig star at a distance of more than 200pc and with a radius of 5 solar radii, then its angular radii will be 0.1mas. For 100m baseline, its visibility will be $V = 0.9986$. To simplify our model, we assume $V_{star}=1$.

For the star we have 3 parameters : the radius ($R_*$), the distance ($d$) and the temperature ($T_*$).
If we asumme a Herbig AeBe star with a temperature of 12000K radiating as a black body, we know that in NIR we will look on the Rayleigh-Jeans regime of a black body (see Fig. \ref{fig:model}). That means that the spectral curve is proportional to a power law : $F_{\lambda}^{star} \propto \lambda^{-4}$. 

\subsubsection{The disk}
The disk model is simple : it is a geometrically thin optically thick passive disk. Its temperature is a function of the radius : 
\begin{equation}
T(r) =  T(r_0) \big( \frac{r}{r_0} \big) ^{-q} 
\end{equation}
with :
\begin{equation}
q = \frac{3}{4} 
	\end{equation}
see references \citenum{Lynden,Adams}.

\begin{figure}[!t]
   \begin{center}
   \begin{tabular}{cc}
   \includegraphics[height=7cm]{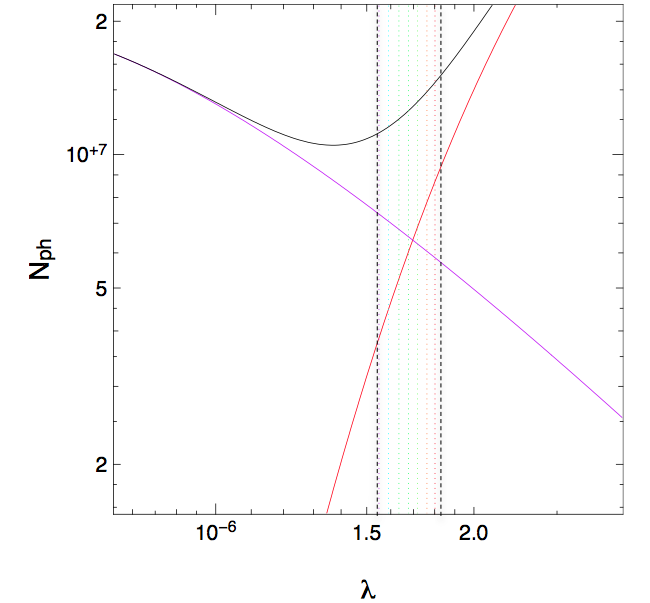}
   \includegraphics[height=7cm]{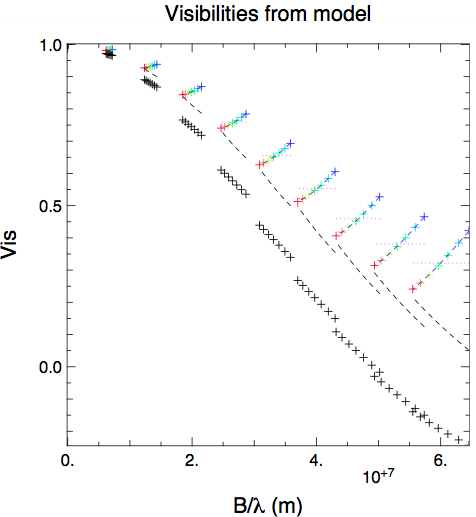}
   \end{tabular}
   \end{center}
   \caption[Results from the disk model] 
   { \label{fig:model} 
On the left, we show the location of the PIONIER spectral channels on the SED of the model. We can see that they are located at the crossing between the stellar and the dust fluxes. On the right, we can see that the chromatic phenomena is reproduced.}
   \end{figure}

The disk will be sampled on several rings, each ring having its own temperature as a function of its distance to the star.
The other geometrical parameters are the inclination ($i$), the inner and outer rims radii ($R_{in}, R_{out}$), and the temperature of the inner rim ($T_{in}$). 
The flux of each ring will be a black body at the temperature of the disk. The ring visibility is defined as follow\cite{Berger} : 
\begin{equation}
\label{eqn:Vring}
V_{ring} = J_0(2 \pi r \frac{B}{\lambda})
\end{equation}
To obtain the Fourier Transform of the disk we have to add the flux of each ring and sum every contributions : 
\begin{equation}
V_{disk} = \sum^{n_{ring}} J_0(2 \pi r_i \frac{B}{\lambda}) 2 \pi r_i B_{\lambda}(T_i) dr_i 
\end{equation}
In the results shown in Fig. \ref{fig:model} the chromatic effect which tends to look like the data shown in Fig. \ref{fig:data}. The visibilities have the same behavior than the data. We can conclude that the chromatic effect is not instrumental but astrophysical. 

The effect is dominated by the flux ratio which is changing through the different spectral channels.
If we compute the total correlated flux we have : 
\begin{equation}
\Vtot(B/\lambda) F_{\mathrm{tot}}(\lambda) = F_*(\lambda) + \Vdisk(B/\lambda) F_{\mathrm{disk}}(\lambda)
\end{equation}
with :
\begin{equation}
F_{\mathrm{tot}}(\lambda) = F_*(\lambda) + F_{\mathrm{disk}}(\lambda)
\end{equation}
If we introduce the stellar to total flux ratio $f_*$, we obtain the mathematical description of the chromatic phenomena :
\begin{equation}
\label{eqn:polyformulae}
\Vtot(B/\lambda)  = f_*(\lambda) + \Vdisk(B/\lambda) (1-f_*(\lambda))
\end{equation}
with
\begin{equation}
f_*(\lambda)  = \frac{F_*(\lambda)}{F_{\mathrm{tot}}(\lambda)}
\end{equation}

In the next section we will discuss the different methods to overcome the chromatic effect.

\section{METHODS} 
\label{methods}

Our goal is to be able to analyze chromatic data. We developed three complementary methods to do that : gray image reconstructions, data modification and parametric fit. The first two methods are based on image reconstruction and the last one is model fitting. We are mostly interested in the disk around the star and we are looking for informations on the resolved geometry and the strength of the chromatic effects.

\subsection{Image reconstruction per spectral channel}

\begin{figure}[ht]
   \begin{center}
   \begin{tabular}{cc}
   \includegraphics[height=7cm]{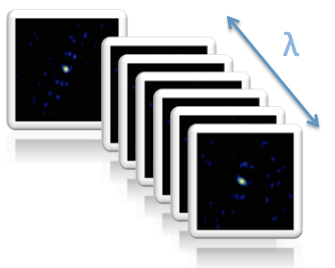}
   \end{tabular}
   \end{center}
   \caption[Results from the disk model] 
   { \label{fig:perlambda} 
   An image reconstruction is made for each spectral channel of the instrument. Then all the images are stacked together in order to obtain the final image.}
   \end{figure}

Once we are aware of the chromatic effect, one can make image reconstructions selecting only one wavelength per reconstruction (see Fig. \ref{fig:perlambda}). In that case the gray image reconstruction is justified. The technique is to have one image per wavelength and to stack all the images in order to have the final broadband one. 

The presence of various components of different spectral indexes prevents from using a gray emission approximation in the image reconstruction  process. As a consequence, since we need to work on a per-spectral-channel basis the (u,v) coverage quality is severely affected.

\subsection{Modification of the data}

We want to have access to the disk visibilities.  From the Eq. (\ref{eqn:polyformulae}),
if we know the SED and then the stellar flux ratio $f_*(\lambda)$ and its variation through the wavelengths, we can compute the disk visibilities as : 
\begin{equation}
\label{eqn:vdisk}
\Vdisk(B/\lambda)  = \frac{\Vtot(B/\lambda) - f_*(\lambda)}{1-f_*(\lambda)} 
\end{equation}

We can apply the modification described in Eq.(\ref{eqn:vdisk}) to one of the interferometric measurements which is the power spectrum ($VV^* = |V|^2$). In summary, our measurements are $|V_{\data}|^2$ and we want to recover $|V_{disk}|^2$. Using the Eq(\ref{eqn:vdisk}), we have : 
\begin{equation}
\label{eqn:v2disk}
\Vdisk^2(B/\lambda)  = \big(\frac{\sqrt{|\Vdata(B/\lambda)|^2} - f_*(\lambda)}{1-f_*(\lambda)}\big)^2 
\end{equation}
One of the problem is the value that we take for $\sqrt{|\Vdata(B/\lambda)|^2}$ ; we must choose between the positive (phase $\phi = 0$) and the negative one ($\phi = \pi$). But it could be solved analyzing more precisely the data and other interferometric observables like the phase of the bispectrum (also called the closure phase).

It is not possible to retrieve the bispectrum phase of the dust from the data because we are loosing the phase of each pair of telescopes (see Eq. \ref{eqn:closurephase}). The equations lead to a solution where we need the phase\cite{Ragland} .

\subsection{Parametric model}
\label{paramfit}

In this section we have attempted to model the object.
The model is geometrical and includes the chromatic effect as described in the sections \ref{sec:intro} and \ref{polychromatism}. Our model is composed of multiple components and was developed when chromatic data was fitted.

\subsubsection{Geometric part of the fit}
The first component of the model is an unresolved star (a dirac in the image space) which can be shifted compared to the image photo center (that will produce a rise of closure phases).
The second component is a ring. In the Fourier space the ring is defined as in Eq. (\ref{eqn:Vring}) but using $\sqrt{u^2+v^2}$ for the spatial frequencies ($B/\lambda$) and their orientations that we want to solve.

In order to be able to have a Position Angle ($P.A.$), which is defined from the North to the East, and a inclination ($i$) we will modify the $uv$-plan for the extended component as follows : 
\begin{equation}
u_{\mathrm{ring}} = u \cos{P.A.} + v \sin{P.A.}\\
\end{equation}
\begin{equation}
v_{\mathrm{ring}} = ( - u \sin{P.A.} + v \cos{P.A.} ) \cos i \\
\end{equation}
One of the parameters of this shape is the ring radius $r$. But this will define a ring with a infinitely small width. In order to have a Gaussian width we have to convolve the ring formulae by a Gaussian, in other words, to multiply the visibility of the ring by the visibility of the Gaussian function with the correspondent width $\mathrm{w}$ :
\begin{equation}
V_{\mathrm{gaussian ring}} = V_{\mathrm{ring}} \, \exp{\frac{-(\pi \mathrm{w} \frac{B}{\lambda})^2}{4 \ln 2}}\\
\end{equation}
Once we have the Gaussian ring, we will add some azimuthal modulations of the ring intensity to be closer to the physics of an inner rim. The modulations are functions in cosinus and sinus of the azimuthal angle ($\alpha$) of the ring which starts at its major axis. We have included two sorts of modulation : one on 2 $\pi$ ($c_1, s_1$) and the second on $\pi$ ($c_2, s_2$).
These modulations are defined as : 
\begin{equation}
I_{\mathrm{tot}} = I_{\mathrm{ring}} (c_1 \cos{\alpha} + s_1 \sin{\alpha} + c_2 \cos{2 \alpha} + s_2 \sin{2 \alpha}) 
\end{equation}
In the visibilities that gives : 
\begin{equation}
V_{\mathrm{tot}} = V_{\mathrm{ring}} - i J_1(2 \pi \frac{B}{\lambda} r) (c_1 \cos{\alpha} + s_1 \sin{\alpha}) - J_2(2 \pi \frac{B}{\lambda} r)(c_2 \cos{2 \alpha} + s_2 \sin{2 \alpha}) 
\end{equation}

The set of parameters defines geometrically a Gaussian ring that can be fitted to the data. 

To complete the model, we can add a second Gaussian ring or a Gaussian function. The Gaussian function is defined with its own position angle ($P.A._{\mathrm{gauss} }$) and inclination ($i_{\mathrm{gauss}}$). Its visibility is defined as follows : 
\begin{equation}
V_{\mathrm{gauss}} = e^{\frac{-(\pi r_{\mathrm{gauss}} \frac{B}{\lambda})^2}{4 \ln 2}}
\end{equation}
with $r_{\mathrm{gauss}}$ being the Half Width at Half Maximum (HWHM) of the Gaussian function.

We have a model with three components : the star, the Gaussian ring and a second Gaussian ring or a Gaussian function.

The total visibilities are depending on all these components weighted by their flux.

\subsubsection{Modeling the chromatism}

To obtain the model visibilities we use the linearity property of the Fourier transform. 
\begin{equation}
\label{eqn:ftot}
F_{\mathrm{tot}} \Vtot = F_* V_* + F_1 V_1 + F_2 V_2
\end{equation}

The fluxes are the ones recieved by the interferometric instrument. So we have : 
\begin{equation}
F = \int_{\lambda - \frac{\Delta \lambda}{2}}^{\lambda + \frac{\Delta \lambda}{2}}F_{\lambda} d\lambda = \int_{\nu - \frac{\Delta \nu}{2}}^{\nu + \frac{\Delta \nu}{2}}F_{\nu}d\nu
\end{equation}
We will use the approximation that the channel spectral width is constant and that the flux is constant in one spectral band. The flux is then equal to the value of $F_{\lambda}$ at the central wavelength of a spectral channel. The exact value of all the terms are described in the Appendix \ref{App}.
From the Eq. (\ref{eqn:ftot}), we see that we can determine a flux ratios at one wavelength and to deduce the ratios on the other wavelength by the laws that we assume for each component.
PIONIER is operating in the NIR in the H band. At this wavelength, we can assume that Herbig stars are in their Rayleigh-Jeans regime. That means that their flux ($F_{\lambda}$)is proportional to the wavelength at the power of $-4$.
The laws for the environment are more difficult to find. We can fit a power-law in wavelength or to a black body variation if we are resolving a thermal emitting region. Since the dust temperature is supposed to be below 2000K\cite{Duschl ,Dullemond} , we can assume that it is in its Wien regime. Then if we assume black body regimes we obtain : 

\begin{equation}
\Vtot(B, \lambda) = f_*^0 \big( \frac{\lambda}{\lambda_0}\big)^{-4} + f_1^0 \frac{B(\lambda, T_1)}{B(\lambda_0, T_1)}  V_1(B, \lambda) + f_2^0 \frac{B(\lambda, T_2)}{B(\lambda_0, T_2)} V_2(B, \lambda)
\end{equation}
with $f^0$ the flux ratios at $\lambda_0$, $T$ the temperature of a component and
\begin{equation}
B(\lambda, T) =\frac{2 h c \lambda^{-5}}{\exp{\frac{h c}{k_B \lambda T}}-1}
\end{equation}
is the black body function, with $h$ the Planck constant, $c$ the light speed, and $k_B$ the Boltzmann constant.

The variations of the flux ratios through the observational band will build the chromatic effect that we want to take into account in our fit.

Once we get all our tools to investigate data with chromatic effect, let us apply them on an astrophysical case : MWC158.

\section{THE CASE OF MWC 158}
\label{mwc158}

The interest on this object came with the data we get with PIONIER\cite{JBLB} a 4 telescopes interferometric, visitor instrument operating at the VLTI and which observe in the H band. 



\subsection{Image reconstructions}
\begin{figure}[!t]
   \begin{center}
   \begin{tabular}{cc}
   \includegraphics[height=7cm]{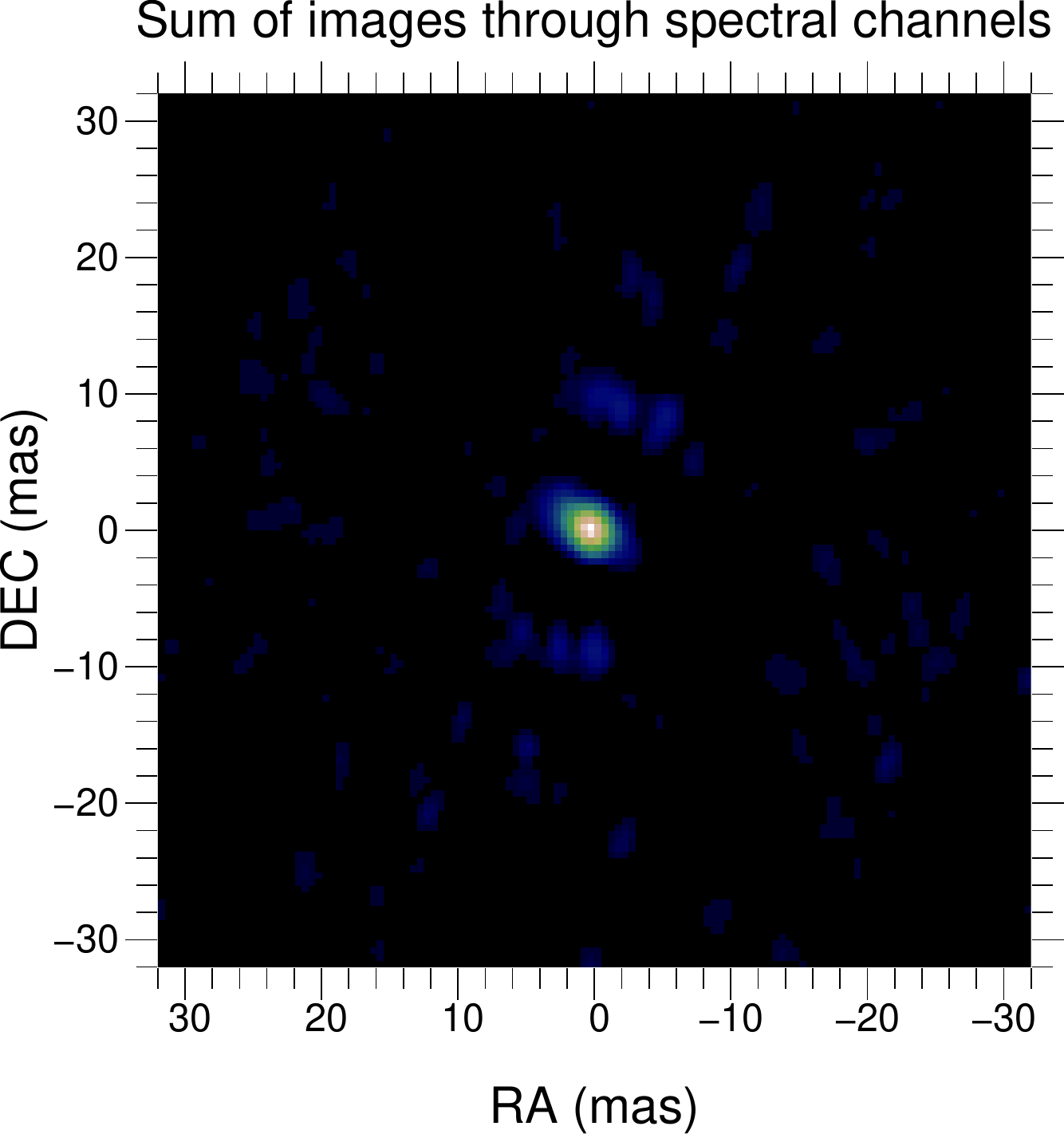}
   \includegraphics[height=6cm]{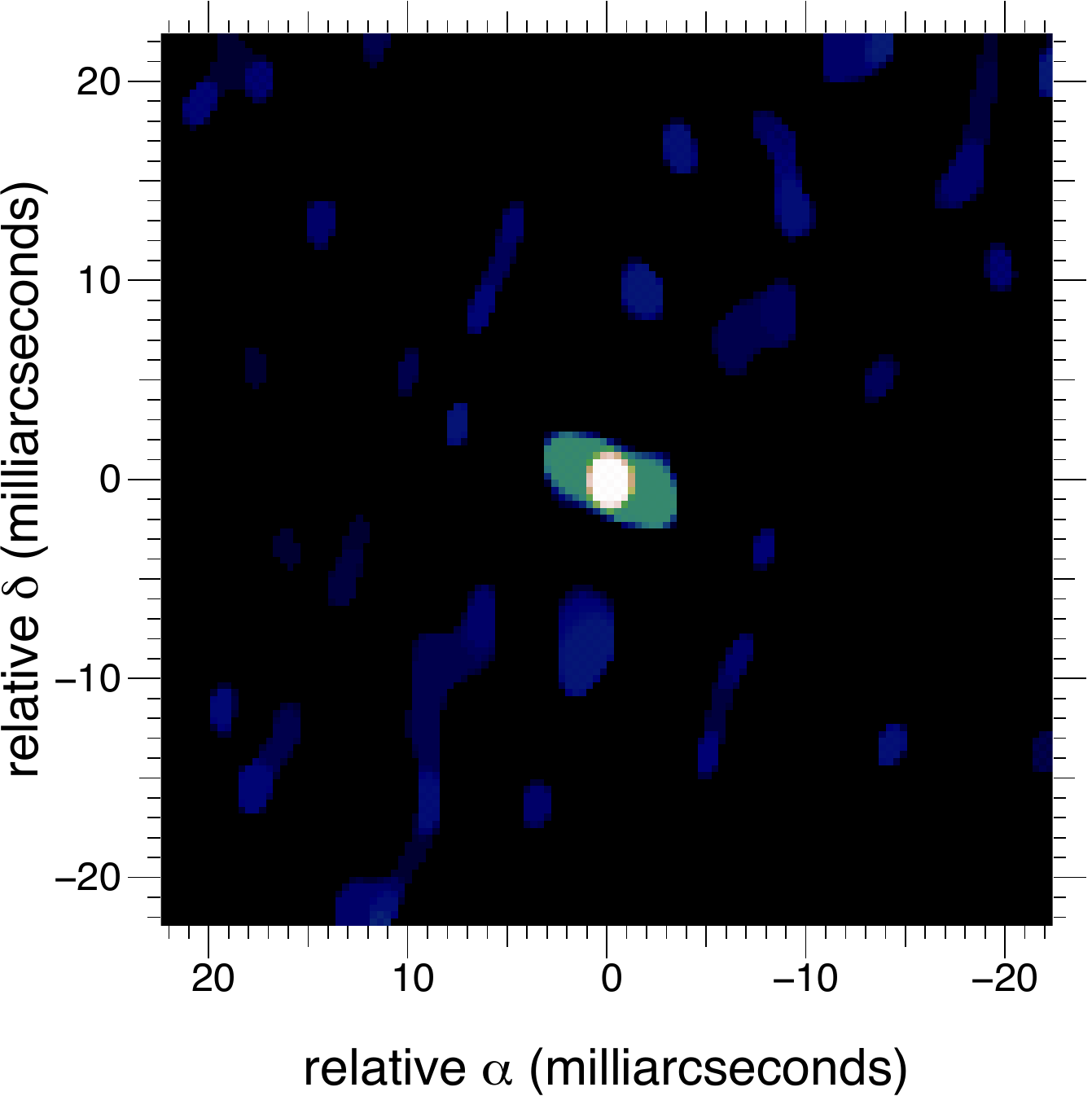}
   \end{tabular}
   \end{center}
   \caption[Image reconstructions on MWC158] 
   { \label{fig:imgrec} 
Left : the stack of  image reconstructions per spectral channels. Right : image reconstruction after modifying the squared visibilities.}
   \end{figure}

We were interested into this data (see Fig. \ref{fig:data}) because it shows clearly signs of chromatism. As the $(u,v)$-plan is sufficiently covered we can reconstruct images. We use the $\tt{Mira}$ algorithm\cite{mira} , but as many image reconstruction algorithms it does not take into account the chromatism. Since it extrapolate the Fourier space, the chromatism makes him extrapolate badly and many artifacts appear.
We then use the monochromatic reconstructions per spectral channel. We also use the visibility correction, not modifying the closure phases (they should be stronger). The results are showed on the Fig. \ref{fig:imgrec}.

We can see that there is a second resolved component. We can also see the orientation of the smallest extended component. 
Both of the reconstruction methods shows similar patterns.
That brings us to the idea to fit two extended components.

\subsection{Parametric fit}
   \begin{figure}[!t]
   \begin{center}
   \begin{tabular}{cc}
   \includegraphics[height=5cm]{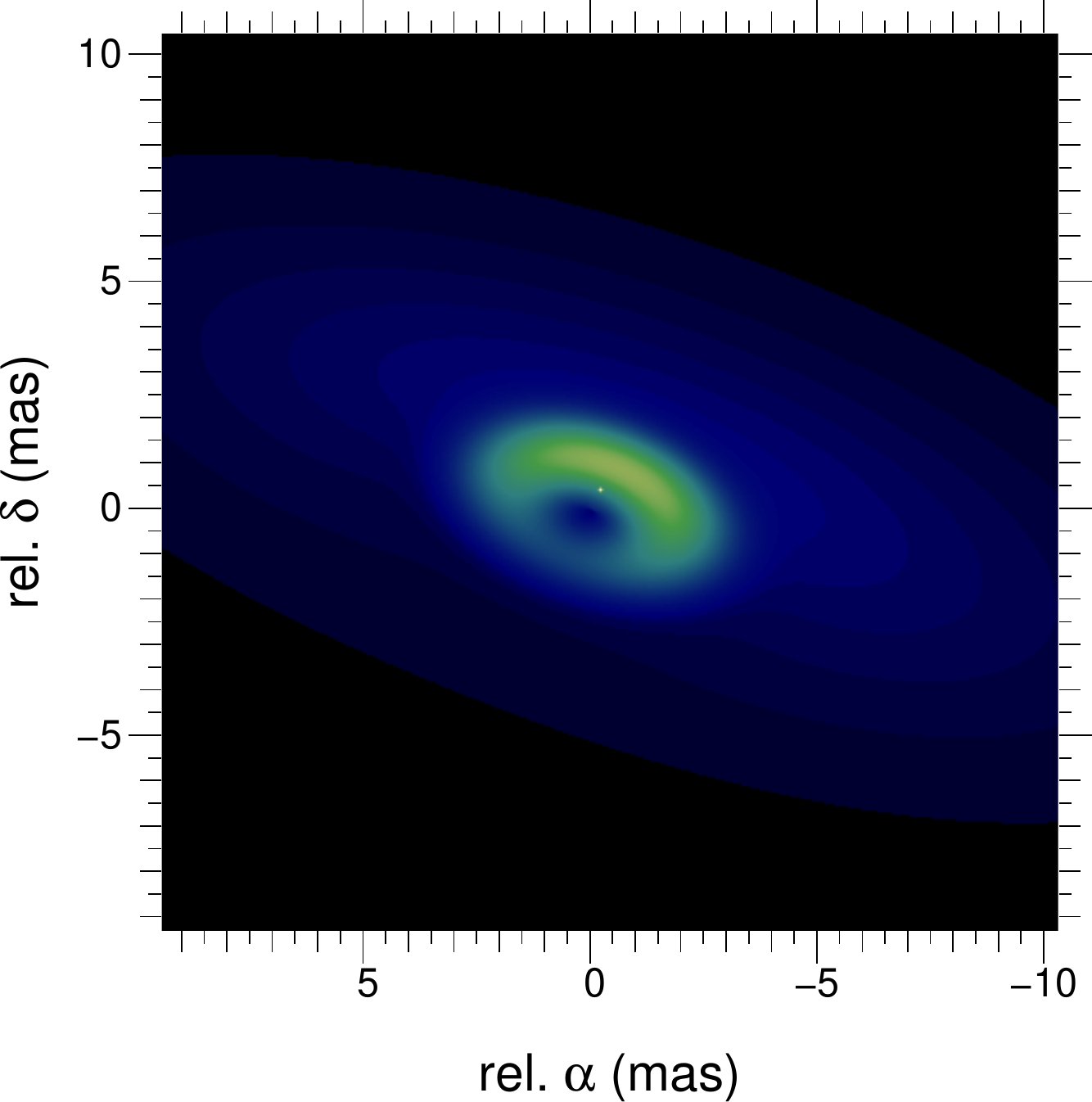}\\
     \includegraphics[height=9cm]{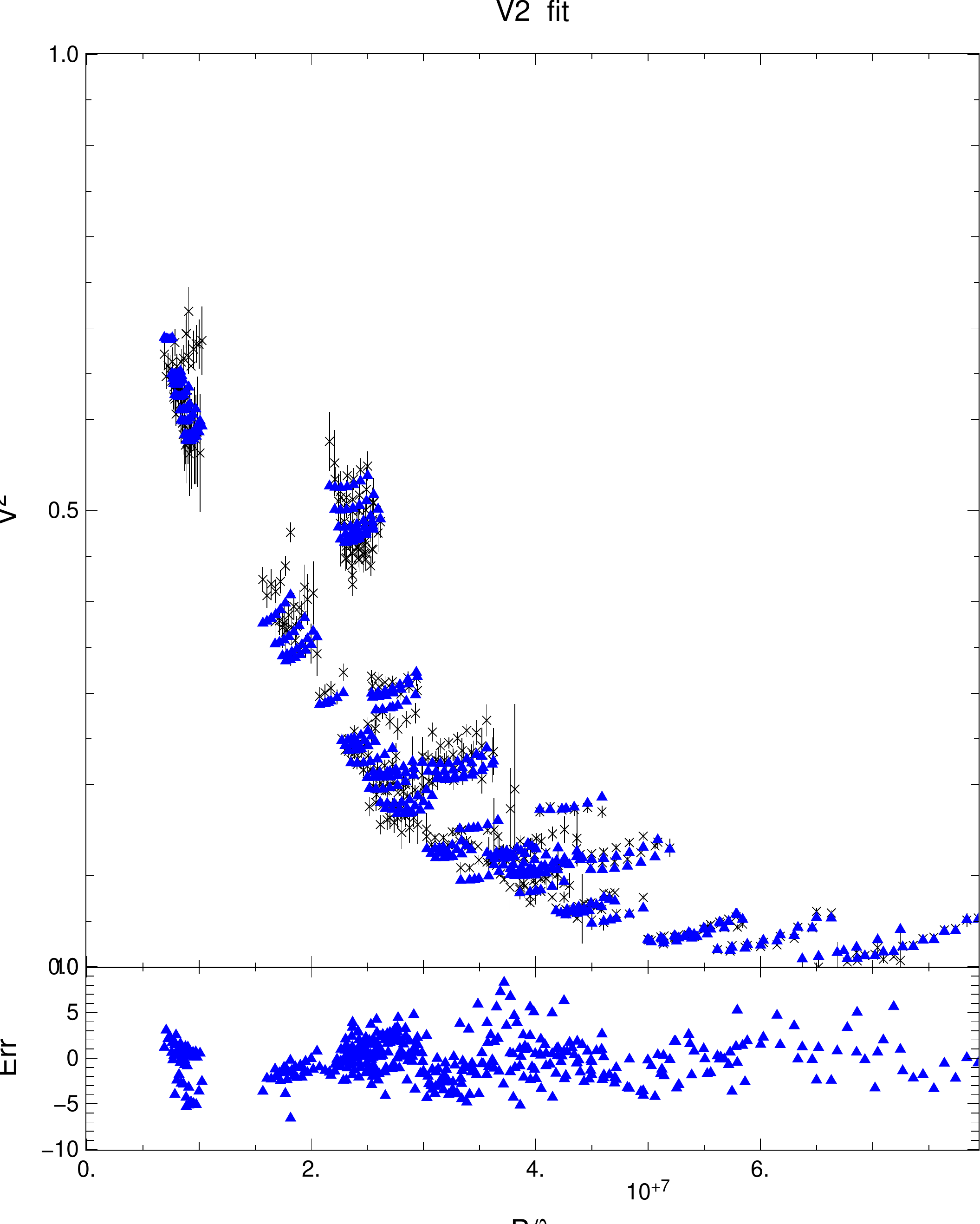}
       \includegraphics[height=9cm]{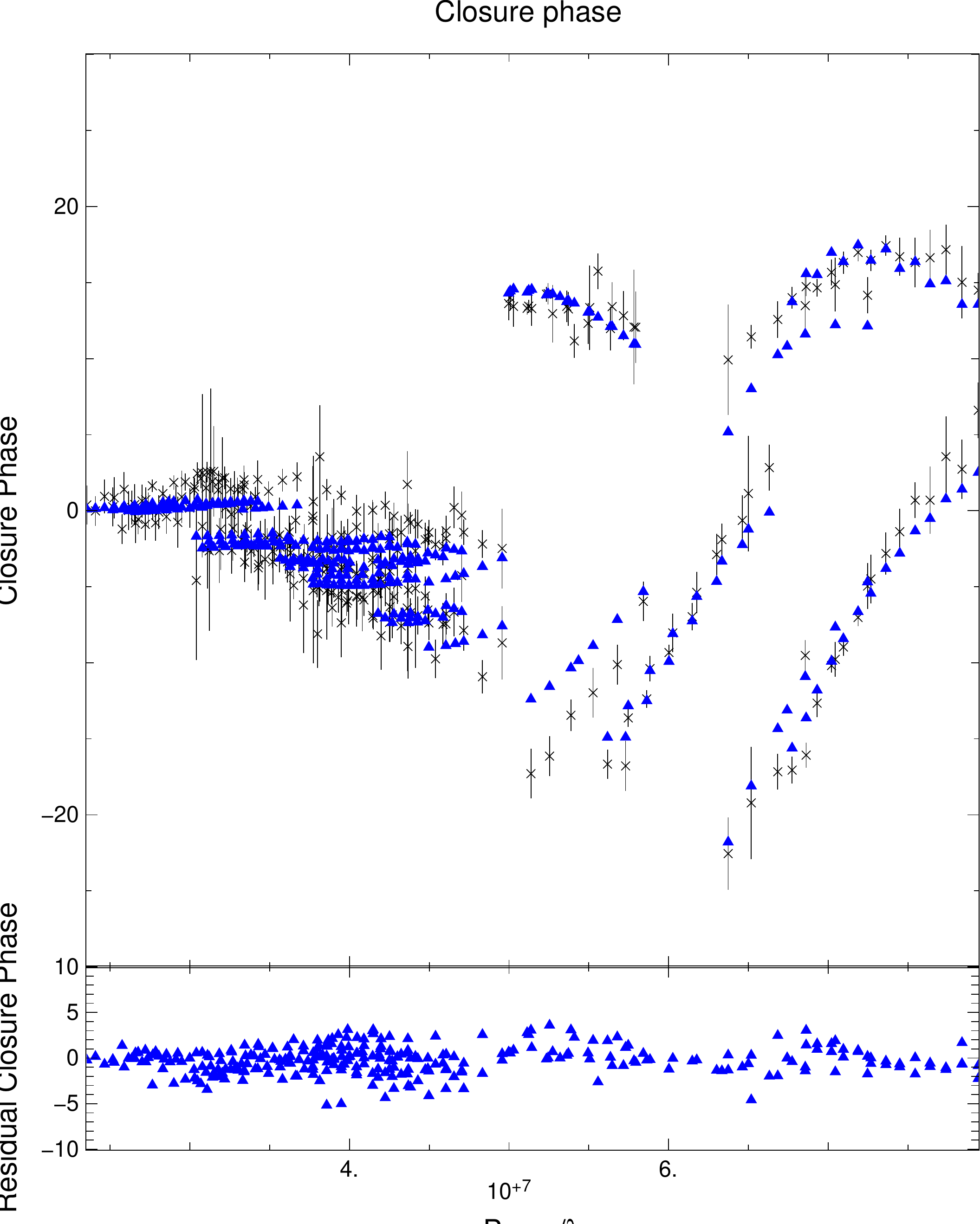}
   \end{tabular}
   \end{center}
   \caption[] 
   { \label{fig:fitmwc158} 
The best fit results are presented. Left : The image corresponding to the best fit. Center : The fit on the V2. Left the fit on the Closure Phases.}
   \end{figure} 

The fit bring us an idea on the geometry and the light emission from the source but with a strong a priori which is the model we want to fit. That is why we took the geometries suggested by the image reconstructions. We can see a central extended part which is composed of the star and its environment which seems to have a P.A. and an inclination.

We have begun the fit with one extended component which is a Gaussian function or a Gaussian ring. Both of the fits gave us the more or less the same inclinations and P.A. which are consistent with the image reconstructions. 
But the data was not entirely fitted : the short baselines indicates that there is a more extended component as showed by image reconstruction. We then add another component to our fit. 
In order to fit the strong closure phase signal we add azimuthal modulation to the ring. It appears not to be sufficient, and the best fit was to shift the central star. It is the only solution to fit the closure phases.

In the end, and adding the different parameters, we ended with 15 parameters and a $\chi^2$ of 3.5.
In the current state of the data processing and interpretation, we believe that the best fit is presented Fig. \ref{fig:fitmwc158}. The parameters are on the Table \ref{tab:fit}. We can see that the best fit is done with two Gaussian rings.

\begin{table}[!t]
\caption{The parametric fit results. The acronyms are described in the section \ref{paramfit}.} 
\label{tab:fit}
\begin{center}       
\begin{tabular}{|l|c|c|l|c|c|l|c|c|} 
\hline
\rule[-1ex]{0pt}{3.5ex}  $\chi^2$  & 3.53 & & & & & & &  \\
\hline
\rule[-1ex]{0pt}{3.5ex}  Star & & & Ring 1 & & & Ring 2 & &  \\
\hline
\rule[-1ex]{0pt}{3.5ex}  Parameter & Value & Error &Param. & Value & Error &Param. & Value & Error  \\
\hline
\rule[-1ex]{0pt}{3.5ex}  $sfr_0$  & 18.0 \% & $\pm$ 0.7\% &$r1fr_0$  & 58.7\%& $\pm$ 1.1\% &$r2fr_0$  & 23.3 \% & $\pm$ 1.8\%   \\
\hline
\rule[-1ex]{0pt}{3.5ex}  &&&$T_1$  & 1482 K & $\pm$ 79 K & $T_2$ & 1326 K & $\pm$ 48 K \\
\hline
\rule[-1ex]{0pt}{3.5ex}  $x_*$  & -0.22 mas  & $\pm$ 0.01 mas & $\mathrm{rr}_1$  & 1.76 mas  &$\pm$ 0.02mas & $\mathrm{rr}_2$  & 3.90 mas & $\pm$ 0.02 mas \\
\hline
\rule[-1ex]{0pt}{3.5ex}  $y_*$  & 0.40 mas & $\pm$ 0.01 mas & $\mathrm{w}_1$  & 1.94 mas & $\pm$ 0.13 mas & $\mathrm{w}_2$  & 13.44 mas & $\pm$ 0.32 mas \\
\hline
\rule[-1ex]{0pt}{3.5ex}  & & & $P.A.$  & 71.67 & $\pm$ 0.64¡ & $P.A.$  & 71.67¡ & $\pm$ 0.64¡ \\
\hline
\rule[-1ex]{0pt}{3.5ex}  &&& $i_1$  & 52.6 & $\pm$ 1.3 & $i_2$  & 67.7 & $\pm$ 1.4  \\
\hline
\rule[-1ex]{0pt}{3.5ex}  &&& $c_1$  & 0.126 & $\pm$ 0.013 &  $c_1$ & 0.126 & $\pm$ 0.013 \\
\hline
\rule[-1ex]{0pt}{3.5ex}  &&&$s_1$  & -0.593 & $\pm$0.033 & $s_1$ & -0.593 & $\pm$ 0.033 \\
\hline
\end{tabular}
\end{center}
\end{table} 

The geometrical fit suggests a star, with a relatively close Gaussian ring (radius of 1.5 mas) with a lot of flux ($\approx$ 60\%). We interpret that as the resolution of the inner rim of the dusty disk. Its azimtuhal modulation is strong in the semi minor axis direction which leads us to deduce that it is due to the inclination. Moreover, the star is shifted towards the most brilliant part of the inner rim. It indicates that the inner rim has a non-negligible height. The outer ring suggests the continuation of the disk, or a part of the disk which is not self shadowed, or a halo. The constrains are poor so we can not conclude on its origins.

The results are shown in Table \ref{tab:ohterinterf}. They are closed to the images get by reconstruction. The results are also consistent with that found with previous observations\cite{Monnier2,Borges2011}  . The authors found similar $P.A.$ with close values of the inclination of the most luminous extended object.

\begin{table}[!t]
\caption{The previous interferometry results on MWC 158. Some results were complete on instruments watching at longer wavelengths (10.7 or 2.2 $\mu m$). The $P.A.$ are consistent and the inclinations $i$ also.} 
\label{tab:ohterinterf}
\begin{center}       
\begin{tabular}{|c|c|c|c|c|c|l|} 
\hline
\rule[-1ex]{0pt}{3.5ex}  FWHM & FWHM2 & $i$ & $P.A.$ & $\chi^2$ & $\lambda_0$ & Ref.  \\
\hline
\rule[-1ex]{0pt}{3.5ex}  66 $ \pm $4 & & 45 & 63 $\pm$ 6 & 1 & 10.7 $\mu m$ & \citenum{Monnier2}   \\
\hline
\rule[-1ex]{0pt}{3.5ex}  64.7 $ \pm $0.6 & & 70.1 $\pm$ 0.7 & 59.1 $\pm$ 1.7 & 5.1 & 10.7 $\mu m$ & \citenum{Borges2011}   \\
\hline 
\rule[-1ex]{0pt}{3.5ex}  35.2 $ \pm $1.5 & 131.4 $ \pm$ 11.2 & 56.7 $\pm$ 0.4 & 65.9 $\pm$ 2.0 & 1.9 & 10.7 $\mu m$ & \citenum{Borges2011}   \\
\hline 
\rule[-1ex]{0pt}{3.5ex}  4.4 $ \pm $0.5 &  & 54 $\pm$ 8 & 66 $\pm$ 9 & 40.8 & 2.2 $\mu m$ & \citenum{Borges2011}  \\
\hline 
\rule[-1ex]{0pt}{3.5ex}  3.0 $ \pm $0.4 & $\geq $14.0 & 54 $\pm$ 8 & 77 $\pm$ 2 & 13.3 & 2.2 $\mu m$ & \citenum{Borges2011}    \\
\hline 
\end{tabular}
\end{center}
\end{table}

The fit of the chromatism, indicates us a black body temperature of the inner rim of $\approx$ 1500K (see Table \ref{tab:fit}). This is approximately the dust sublimation temperature found in the litterature\cite{Duschl,Dullemond} .

\section{Conclusion and Perspective}

The chromatic effect due to the flux predominances of two objects of different sizes is well understood and can be used in order to find astrophysical information of the object. In the case of Herbig AeBe stars we are able to have an approximation of the temperature of the components. If the chromatic information is given, we can perform gray disk image reconstructions. They are contributing to the astrophysical analyze of the object because they shows the P.A. and the inclinations of the disk. Moreover, in the case of MWC158 it brought us the idea of the second extended component. By the fit we were able to find the inner rim radius and its temperature and to compare what we found with the data from photometry. We bring the first confirmation of the dust sublimation temperature at the inner rim. The information taken from the NIR interferometry and the chromatic effect argue in favor of a young nature of MWC 158.

\begin{figure}[!t]
   \begin{center}
   \begin{tabular}{cc}
   \includegraphics[height=7cm]{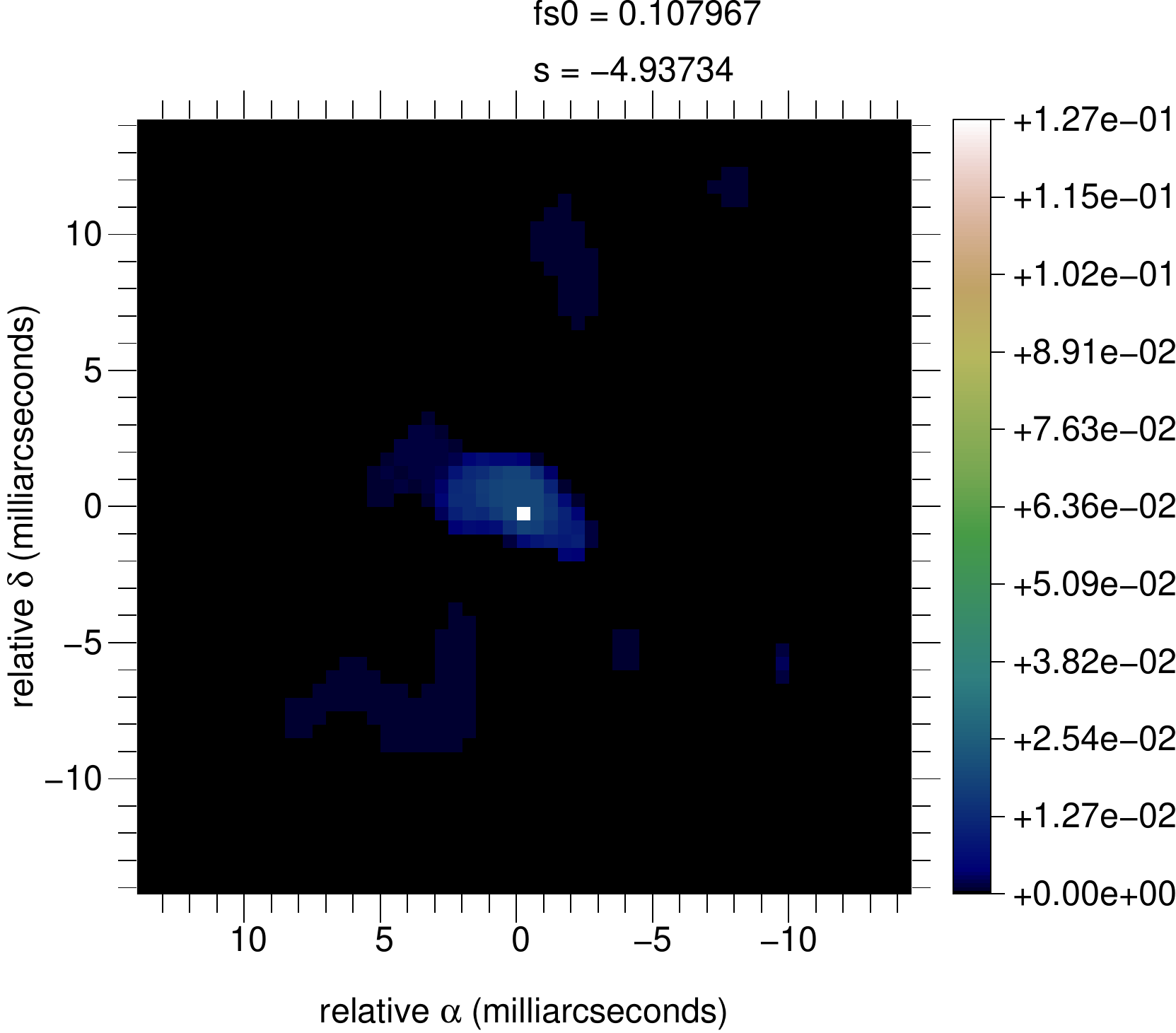}
   \end{tabular}
   \end{center}
   \caption[] 
   { \label{fig:ImgRecPM} 
Image corresponding to a polychromatic reconstruction with the $\tt{mira}$ algorithm.}
   \end{figure}

The main challenge is to be able to make chromatic Young Stellar Objects image reconstructions keeping the super spectral synthesis and without information on the total flux variation. One of the thing which is in process of testing, is the adaptation of the $\tt{Mira}$ algorithm to the case of young stellar object. The "gray" $\tt{mira}$ free parameters are the image pixels intensities. If we define the image as the image of the dust at $\lambda_0$, the start can be represented by a dirac at the center of the image. Hence, we can put the stellar flux and a stellar relative spectral power law as additional parameters to the fit. Since the regularization will tend to smooth the Fourier plan, the algorithm will favors the added parameters to fit the fixture. The first tests was realized on MWC 158 (Fig. \ref{fig:ImgRecPM}). The result found a star with a flux ratio at $\lambda_0$ of 10.7\% and a relative spectral slope of $\kappa_* = -4.93$, which corresponds to a black body temperature the disk image at $1.65 \mu m $ of 900K.

\appendix

\begin{appendix}

\section{Analytic flux formulae}
\label{App}

In this appendix we will present the exact equations for the flux received by each spectral channel in order to be able to compute the good model of flux both for the disk and the star.

The flux received by the instrument is : 

\begin{equation}
F = \int_{\lambda - \frac{\Delta \lambda}{2}}^{\lambda + \frac{\Delta \lambda}{2}}F_{\lambda} d\lambda
\end{equation}

with $F$ is the integrated flux received by the instrument, $\lambda$ the wavelength.

\subsection{The star}

For the star we are in the Rayleigh-Jeans side of the black body flux curve, so we have : 

\begin{equation}
F_{*} = \int \bornes 2 k_B c T \lambda^{-4} d\lambda
\end{equation}

with $k_B$ the Boltzmann constant, $c$ the light speed, $T$ the temperature of the star.

Since we are interested only in the evolution of the fluxes in through the spectral channels we need only : 

\begin{equation}
F_{*} \propto \int \bornes \lambda^{-4} d\lambda = \left[ - \frac{\lambda^{-3}}{3} \right]\bornes
\end{equation}

\subsection{The disk}

For disk will be in the Wien regime of the blackbody curve. So we will have : 

\begin{equation}
F_{disk} = \int \bornes 2 h c^2  \lambda^{-5} \exp{(-\frac{h c}{k_B \lambda T})} ~d\lambda
\end{equation}

with $h$ the Planck constant.

Without the constant parts it would be : 
\begin{equation}
F_{disk} \propto \int \bornes \lambda^{-5} \exp{(-\frac{C_1}{\lambda})} ~d\lambda
\end{equation}

with $ C_1 = \frac{h c}{k_B T}$. In order to have the analytical solution of it, we have to integrate it by parts.
The first step is : 

\begin{equation}
F_{disk} \propto \frac{1}{C_1} \big( \int \bornes 3 \lambda^{-4} \exp{(-\frac{C_1}{\lambda})} ~d\lambda + A \big)
\end{equation}
with $A = \left[ \lambda^{-3} \exp (-\frac{C_1}{\lambda}) \right] \bornes $.

Then : 

\begin{equation}
\int \bornes 3 \lambda^{-4} \exp{(-\frac{C_1}{\lambda})} ~d\lambda =\frac{3}{C_1} \big( 2 \int \bornes \lambda^{-3} \exp{(-\frac{C_1}{\lambda})} ~d\lambda + B \big)
\end{equation}
with $B =  \left[ \lambda^{-2} \exp (-\frac{C_1}{\lambda}) \right] \bornes$.

If we compute : 

\begin{equation}
\int \bornes \lambda^{-3} \exp{(-\frac{C_1}{\lambda})} ~d\lambda = \frac{1}{C_1} \bigg(  C + \int \bornes \lambda^{-2} \exp \big( - \frac{C_1}{\lambda} \big)   \bigg)
\end{equation}

with $C = \left[ \lambda^{-1} exp (- \frac{C_1}{\lambda}) \right] \bornes$.

Finally :

\begin{equation}
\int \bornes \lambda^{-2} \exp{(-\frac{C_1}{\lambda})} ~d\lambda = \frac{1}{C_1} \left[  \exp (- \frac{C_1}{\lambda})  \right] \bornes
\end{equation}

With all these computations the final result for the disk flux is : 

\begin{equation}
F_{disk} = \frac {2 h c^2}{C_1} \Bigg( A + \frac{3}{C_1} \bigg( B +  \frac{2}{C_1} \Big(  C +  \frac{1}{C_1} \left[  \exp (- \frac{C_1}{\lambda})  \right] \bornes \Big)    \bigg) \Bigg) 
\end{equation}

\end{appendix}

\acknowledgments

We would like to thank the A.N.R. POLCA which is funding this study.

\bibliography{report}   
\bibliographystyle{spiebib}   

\end{document}